# Levels of Product Differentiation in the Global Mobile Phones Market

## The sixth product level called compliant product is a connecting element between the physical product characteristics and the strategy of the producer company

**Stanimir Andonov***

The article discusses the differentiation among the product offers of companies working in the global markets, as well as the strategies which they use and could use in that respect.

The main idea of the paper is that the principle “differentiate or die” (Jack Trout) has died. Today the global brands don’t strive to differ from their competitors in everything and at any cost. As an example, let’s have a global look at the business of mobile phones. In June 1998 Ericsson, Nokia, Motorola and Psion established their own International Strategic Alliance, a private independent company called “Symbian”. Symbian Ltd. is an independent, for-profit company whose mission is to establish Symbian OS as the world standard for mobile digital data systems, primarily for use in cellular telecoms. So, the three biggest mobile phone companies at that time and Psion, a mobile PC company, created a new way for competing and differentiating. Today Symbian Ltd. is owned by Ericsson, Nokia, Panasonic, Samsung, Siemens and Sony Ericsson. Figure 1 shows the percentage of shares of each shareholding company:

**Figure 1 Shareholding companies of Symbian and their percentage of shares**

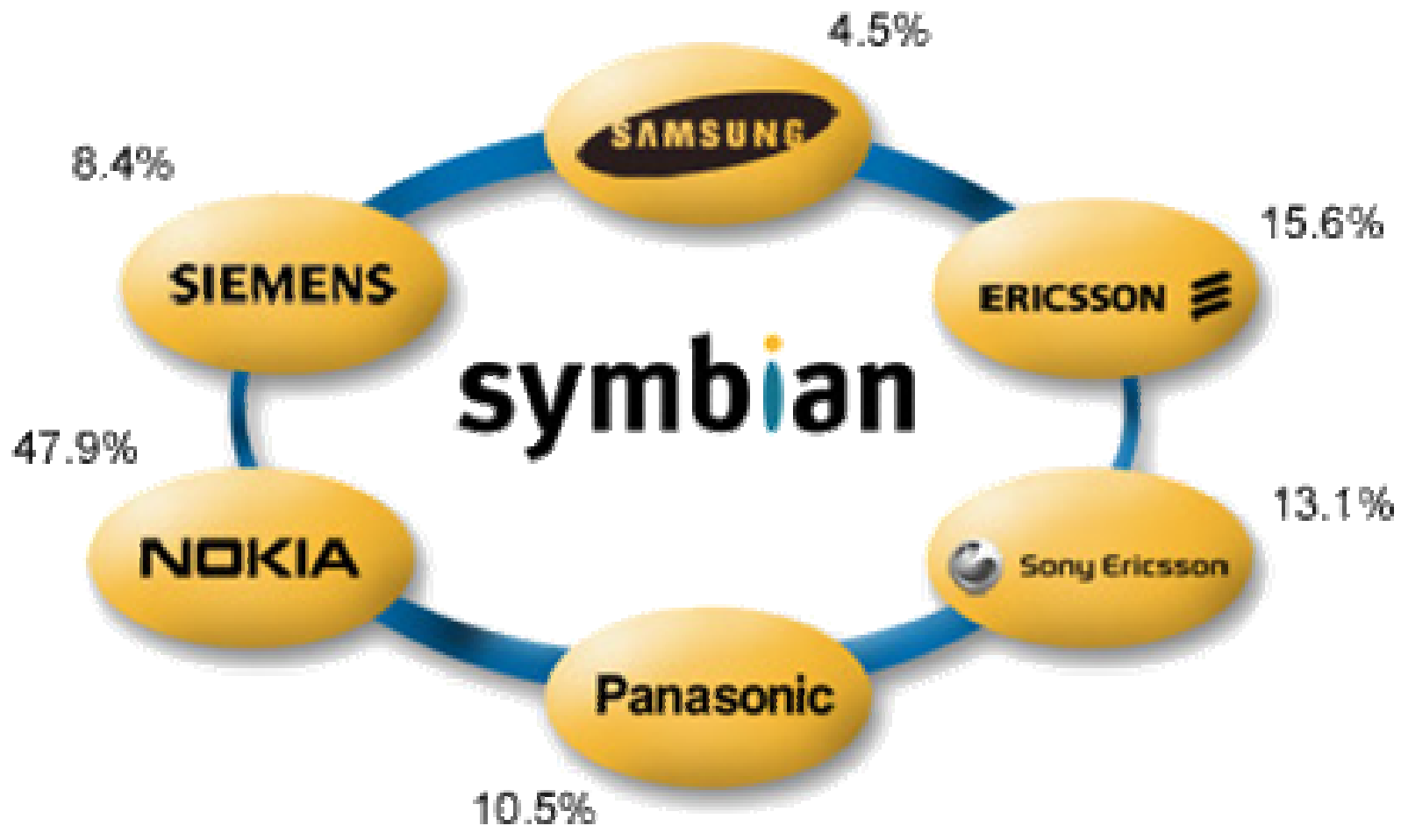

*Source:* Symbian Ltd. – Company Ownership, http://www.symbian.com/about/ownership.html

### Levels of product differentiation

Porter[1] has suggested that most industries contain *strategic groups* of close competitors – groups of firms within an industry that follow the same strategies or ones that have very similar dimensions. For example, similar strategies might include the targeting of the same market segments, the use of identical or similar technology and the employment of the same specialist distributors.

[1] Porter, M., “Competitive Strategy – Techniques for Analyzing Industries and Competitors,” (New York: The Free Press, 1980).

Each global company participates in strategic groups and each group has its conditions of participation which the company has to follow. For example, Nokia participates in the Symbian strategic group with its Smartphones (Nokia 3/6/7/9/N Series), but it participates in other groups with the rest of its series (Nokia 1/2/5/8 and Nokia Vertu).

Theodore Levitt's[2] classic concept of product levels is used to determine the criteria under which companies can be organized into strategic groups. According to him the product levels are four, but later on Philip Kotler[3] adds one more and thus the product levels become five:

1. **Core benefit – Kotler**
2. **The generic product – Levitt**
3. **The expected product – Levitt**
4. **The augmented product – Levitt**
5. **The potential product – Levitt**

This analysis shows that companies compete actively at all these levels (except the first one – core benefit). Competition and differentiation are mainly at the time of market launch.

**Participating in strategic groups is the only strategic competitive advantage which is hard to diminish by time.**

If we assume that the owners of Symbian Ltd. and the Symbian OS licensees (See Figure 2) form a strategic group according to Porter's criteria, the total market share of this strategic group is 79,4% in 2005. Nokia rules the mobile market with a 32.5% share, followed by Motorola (17.7%), Samsung (12.7%), LG (6.7%), Sony Ericsson (6.3%), and Siemens (3.5%)[4].

**Figure 2 Symbian OS Owners and Licensees**

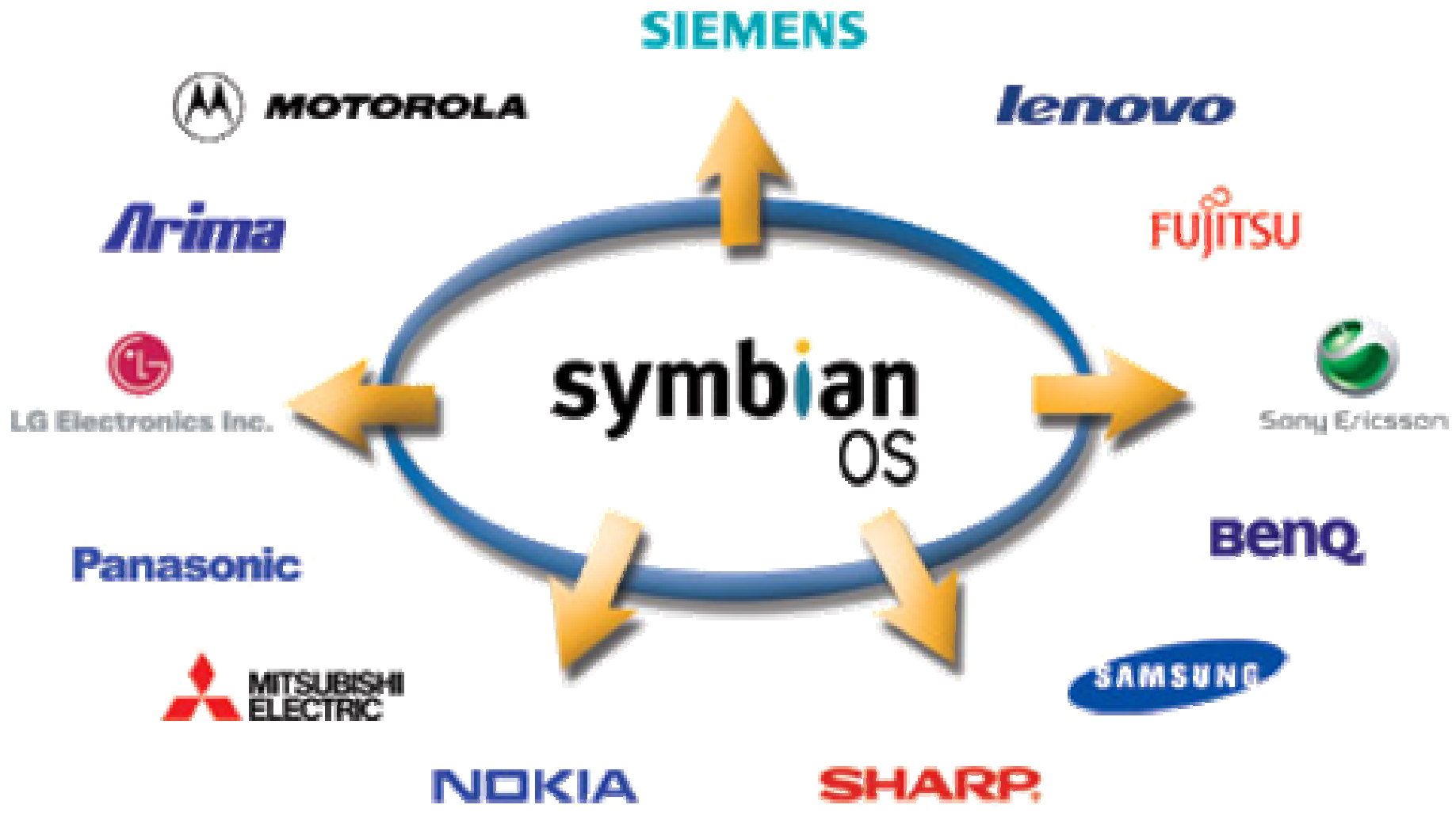

*Source:* Mobile Phone Manufacturers licensing Symbian OS, http://www.symbian.com/about/licensees.html

The Symbian OS licensees make equal use of the advantages that the operating system gives them. On the basis of these criteria we can assume that the products of the companies comprising this strategic group have one product level common for all – the sixth product level.[5] The sixth product level is the level at which they don't compete. The companies have associated in the International Strategic Alliance to create a common competitive advantage. The success of this competitive advantage has later led to the creation of a strategic group through licensing [6](Symbian Ltd. sells licenses).

---

[2] Levitt, Th., "Marketing Success through Differentiation – of Anything" Harvard Business Review 58 (January-February 1980): 83-91.
[3] Kotler, Ph., "Marketing Management," 8th ed. (New Jersey: Prentice-Hall Inc, 1994).
[4] Clarke, P., "Mobile Phone Sales Up 21% In 2005: Gartner", Feb 28, 2006, www.informationweek.com
[5] For the purpose of clarity, I have to specify that these companies use Symbian OS only for their Smartphone models. We can assume that the Smartphone is a GSM device that has an operating system installed or can have one installed together with all advantages related to it.

### The sixth product level – level of the company's compliance with the standard of the strategic group

When everybody in the group has the same product level, this level becomes a level of the company's compliance with the strategic group. For example, a necessary condition of entry into the Symbian OS group is the purchase of a license. In this way the sixth level reduces the differences among the products of the companies in the strategic group and becomes a necessary condition for having the right to be positioned in the market as a part of this group. For instance, all off-road vehicles use almost the same four-wheel-drive system but each brand has a different name for it: *Jeep* – 4x4, *Audi* – Quatro, *Honda* – 4WD, *Renault* – Qadra, *Mercedes* – 4matic, *Toyota* – AWD, etc.

**Figure 3 The Six Product Levels**

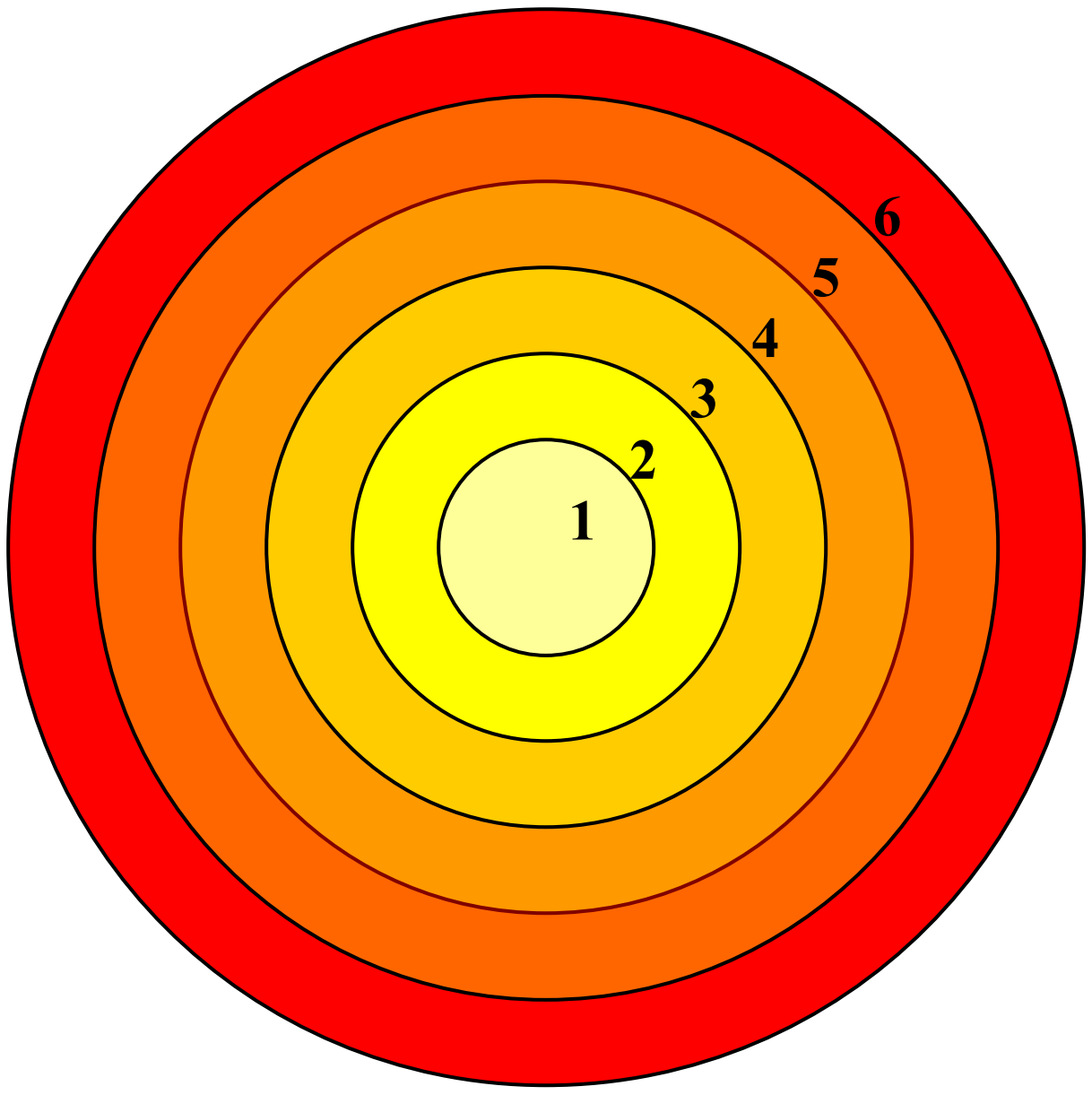

**1. Core benefit**
**2. The generic product**
**3. The expected product**
**4. The augmented product**
**5. The potential product**
**6. The compliant product**

The sixth product level is the level of compliance of the product with the group's standard. We can assume that the sixth product level is a combination of the benefits and functions of the product which are at the other five levels. Thus, $1^{st}$ level + $2^{nd}$ level + $3^{rd}$ level + $4^{th}$ level + $5^{th}$ level = $6^{th}$ level. The sixth level presents the whole way of functioning of the product, the total amount of benefits that it offers to the consumer. For example, all Smartphones today offer to the consumer almost the same array of benefits such as a hard disc memory and a possibility to put an additional memory card, Bluetooth, Infrared, GPRS, EDGE, WAP, built-in modem, WiFi, HSCSD, XHTML, USB, Push-to-talk, Mobile Java – all of them work together thanks to the operating system and represent the functional value of the sixth level. The companies in the International Strategic Alliance Symbian Ltd. create a common competitive advantage which they share with the companies that have purchased a license in order to become a part of the Symbian OS strategic group. They use the operating system in the same way thus avoiding competition among themselves at this level. However, they keep competing at the other levels.

[6] See: Kline, D., "Sharing the Corporate Crown Jewels", MIT Sloan Management Review 44, no.3 (Spring 2003): 89-93.

**The functional value of the sixth level**

The sixth level not only unites the companies in strategic groups thus differentiating whole strategic groups of companies and products, but also has a functional value, as shown in the example. The functional value of Symbian OS can be referred to as *mass customization*. A smartphone that works with Symbian OS gives freedom to its user. The consumers can now themselves choose for what else to use their devices and what software to install on their phones.[7] Mass customization occurs. The concept “mass customization” was mentioned for the first time by Stan Davis in his book “Future Perfect” [Reading, MA: Addison-Wesley] published in 1987. B. Joseph Pine II, in his classic work Mass Customization [Boston: Harvard Business School Press, 1992] develops this concept further and argues that ultimately segmentation may be unimportant because individual customers’ needs might be addressed individually. In this connection, the sixth level and Symbian OS give the consumers the opportunity to customize their mobile phones and use them to their liking. The business philosophy that the companies follow is: *we reduce the opportunities for differentiation among us, but we increase the freedom of our consumers by giving them a well-working, constantly-developing operating system* (you can refer back to Symbian Ltd.’s mission at the beginning of the article). After all, Symbian OS is already a standard and Symbian Ltd.’s mission is about to be accomplished. See figure 4.

On the basis of these facts and analysis we can conclude that the sixth product level can be also called *future product*. Let’s say that you buy Nokia N70. At the moment of purchase you know all technological characteristics of the product and all the options of the operating system (Symbian v8.1), but you don’t know the future application of the mobile phone. Why is that? Because a lot of compatible programmes, games, etc. will be developed in the future for Symbian OS v8.1. Thus you also buy the promise for future support services without knowing what they will be. This gives every consumer the opportunity to be unique.

**Figure 4**

**Worldwide total smart mobile device market**
**Market shares Q2 2005, Q2 2004**

| OS vendor | Q2 2005 shipments | % share | Q2 2004 shipments | % share | Growth Q2'05/Q2'04 |
|---|---|---|---|---|---|
| **Total** | **12,185,600** | **100.0%** | **5,933,330** | **100.0%** | **105.4%** |
| **Symbian** | 7,648,920 | 62.8% | 2,429,930 | 41.0% | 214.8% |
| **Microsoft** | 1,931,630 | 15.9% | 1,360,220 | 22.9% | 42.0% |
| **PalmSource** | 1,157,720 | 9.5% | 1,335,810 | 22.5% | -13.3% |
| **Others** | 1,447,330 | 11.9% | 807,370 | 13.6% | 79.3% |

**Source: Canalys estimates, © canalys.com ltd. 2004-2005**
Smart mobile device market: handhelds, wireless handhelds, smart phones

[7] Apart from Symbian Ltd. a lot of other software companies develop software for the different versions of the operating system worldwide. These programmes can be bought or downloaded for free from the Internet by the consumer. Thus Symbian gives the freedom for those who want to develop software for it and increases the opportunities for application of the phones that have OS.

The sixth product level for IT devices working with OS[8] (GSM, PDA, PC, etc.) is every future development of the product software (installation of different programmes or their update) which will lead to additional benefits for the consumer that he doesn't know about or doesn't expect at the time of purchase.

## Levels of differentiations and competitions among companies and their products

The competition among companies and their products is closely linked to the ways of differentiation among them. The process of successful competitive strategy development and execution starts with the identification of the company's competitors. It continues with the identification and realization of the levels of differentiation of products which satisfy the same needs and wants. The different levels of differentiation among products involve discussion of the different levels of competition among them as part of a company's common competitive strategy. In this article I analyze how participation of companies in strategic groups affects the differentiation of their products. The differentiation of these products can be conceived of in terms of five levels (see Figure 6):

1. Quaternary competition. Among different products which satisfy the same need or want. Product category vs. product category (GSM vs. TDMA; CDMA; DoCoMo; NAMPS; iDEN/Nextel; NMT; TETRA/Dolphin; Iridium; Globalstar).
2. Tertiary competition. Product line vs. product line of the same product category. GSM Smartphones vs. GSM without an operating system.
3. Secondary competition. Product type vs. product type of the same product line. Smartphone with Symbian OS vs. Smartphone with other OS.
4. Primary competition. Competition among the brands of the same product type but with different levels of involvement in the strategic group. Smartphone Nokia with Symbian OS (International Strategic Alliance Symbian Ltd.) vs. Smartphone with Symbian OS (licensee member of Symbian OS strategic group). Sub-strategic group vs. sub-strategic group.
5. Zero competition. Level of competition among the brands of the same product type and of the same sub-strategic group.

[8] Operating systems perform basic tasks, such as recognizing input from the keyboard, sending output to the display screen, keeping track of files and directories on the disk, and controlling peripheral devices such as disk drives and printers. Operating systems provide a software platform on top of which other programs, called *application programs,* can run. The application programs must be written to run on top of a particular operating system.

*Source*: http://www.webopedia.com/TERM/o/operating_system.html

**Figure 5 Levels of competition and differentiation**

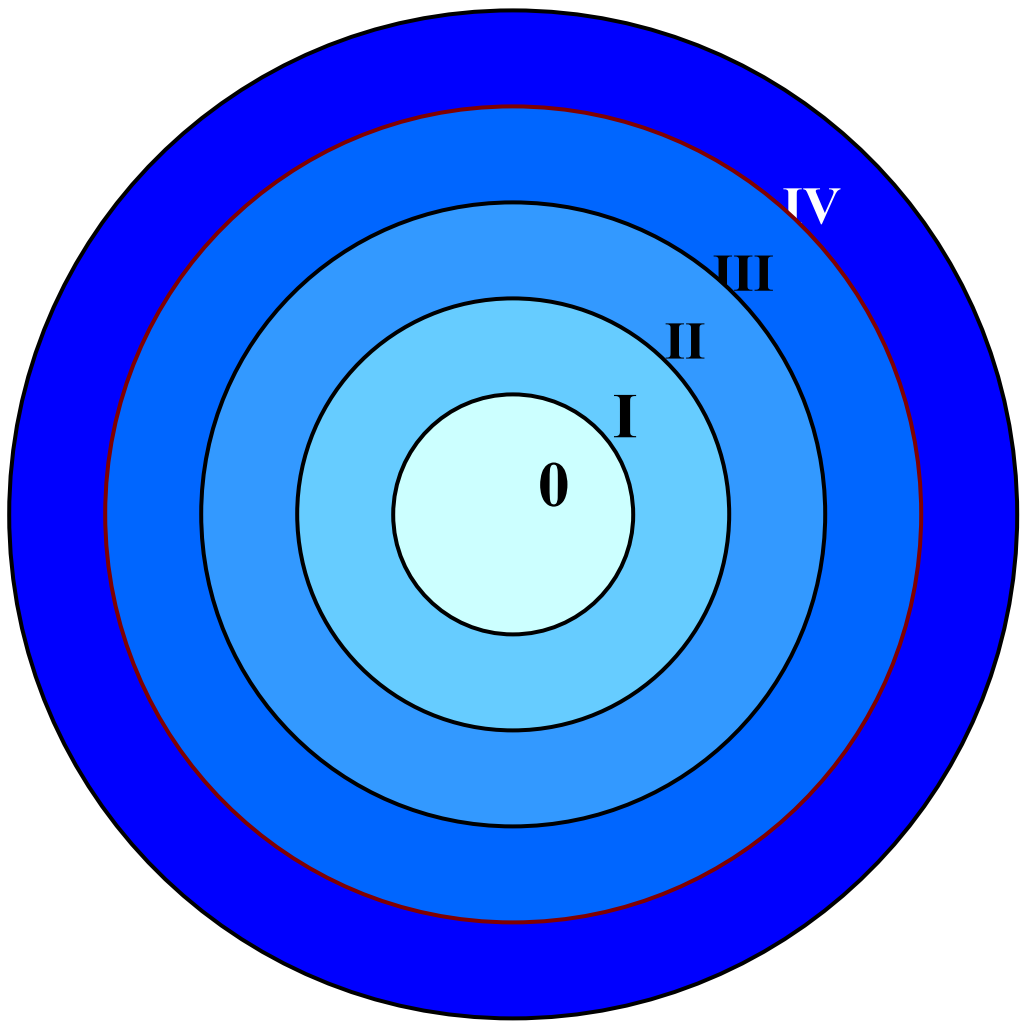

**0. Zero competition**
**I. Primary competition**
**II. Secondary competition**
**III. Tertiary competition**
**IV. Quaternary competition**

As there is no common system for Product hierarchy in the world economic literature, product hierarchies extend from basic needs to specific things that satisfy these needs. I am going to use Professor Ph. Kotler's system of product hierarchy which he proposes in the 8th edition of his book Marketing Management [Prentice-Hall, 1994]. This system is adapted to the object of study:

1. Need family. Information exchange. Communications.
2. Product family. Devices for electronic information exchange. Wireless information devices.
3. Product category (class). GSM devices.
4. Product line. GSM Smartphones.
5. Product type. GSM Smartphones with Symbian OS.
6. Brand. Nokia.
7. Item. Nokia N70.

Here we have to make it clear that the strategic groups of companies can be formed in three different ways and are three different types respectively (see Figure 6).

1. Strategic groups of 1st type.
   1.1 Strategic groups formed by companies through participation in joint ventures (International Strategic Alliances, Symbian Ltd.).
   1.2 Strategic groups formed through purchase of license, purchase of the right for participation in the group (Symbian OS Licensees).
2. Strategic groups of 2nd type. A strategic group consists of those rival firms with similar competitive approaches and positions in the market.[9]
3. Strategic groups of 3rd type. Strategic groups formed by necessity. These groups consist of companies whose products don't comply with the standard of the 6th level. They form not only separate strategic groups but also their own product lines to which they belong because of the incompliance of their products with the 6th level of the products of their competitors.

[9] Porter, M., "Competitive Strategy – Techniques for Analyzing Industries and Competitors," (New York: The Free Press, 1980), pp. 129-30.

**Figure 6 Product hierarchy and levels of competition among products**

| Product categories (classes) Mobile phones / Product line and type | 1. GSM device | 2. TDMA; 3. CDMA; 4. DoCoMo;5. NAMPS; 6. iDEN/Nextel; 7. NMT; 8. TETRA/Dolphin; 9. Iridium; 10. Globalstar). |
|---|---|---|
| | **Quaternary competition** | |
| | The presence of OS in a GSM device is the 6th level | |

| Product line | Product type | Secondary / Tertiary competition | Strategic group | Competition level / subgroup | OS | Members |
|---|---|---|---|---|---|---|
| Product line Smartphones | *Product type* GSM Smartphones with Symbian OS | **Secondary competition**, **Tertiary competition** | Strategic group – 1 type | **Primary competition**, **Zero** – subgroup | | 1.1 ISA Symbian Ltd. Nokia, Ericsson, Sony Ericsson, Panasonic, Siemens, Samsung |
| Product line Smartphones | *Product type* GSM Smartphones with Symbian OS | **Secondary competition**, **Tertiary competition** | Strategic group – 1 type | **Primary competition** – subgroup | | 1.1 OS Symbian – licensees Motorola, Lenovo, Fujitsu, BenQ, Sharp, Mitsubishi electric, LG Electronics Inc., Arima |
| Product line Smartphones | *Product type* Smartphones with other operating systems | **Secondary competition**, **Tertiary competition** | Strategic group – 2 type | A strategic group formed on the basis of one common strategic advantage (6th level) | OS | GSM device |
| Product line Smartphones | *Product type* Smartphones with other operating systems | **Secondary competition**, **Tertiary competition** | Strategic group – 2 type | | Microsoft | Asus, Compal , Microsoft prototype: Firefly, Microsoft prototype: Avenger, Mitac Mio, , Motorola , Orange SPV, Samsung , Sierra Wireless Voq, Tanager by HTC, |
| Product line Smartphones | *Product type* Smartphones with other operating systems | **Secondary competition**, **Tertiary competition** | Strategic group – 2 type | | RIM | RIM BlackBerry |
| Product line Smartphones | *Product type* Smartphones with other operating systems | **Secondary competition**, **Tertiary competition** | Strategic group – 2 type | | Linux | Accton, Cellon, Philips, Datang, FMC, Ericsson, G-Tek, Haier, HTC, ImCoSys, Longcheer/Oswin, Motorola, NEC, Panasonic, ROAD, Samsung, Siemens, SK Telecom, TCL, Telepong, Yahua TelTech, Yulong, Winston NeWeb, ZTE |
| Product line Smartphones | *Product type* Smartphones with other operating systems | **Secondary competition**, **Tertiary competition** | Strategic group – 2 type | | PalmSource | GSPDA, Kyocera, Palm, PiTech, Samsung |
| Product line non-Smartphones | GSM device without OS | **Tertiary competition** | Strategic group – 3 | Strategic group of necessity, formed on the basis of the presence of the 6th level. | | GSM without OS. Alcatel, Sagem |

As it can be seen in Figure 5, the company's compliance with a certain strategic group is manifested by the product's attributes present at the 6th product level. The 6th product level is a criterion which divides the companies' products into lines – each product line is a strategic group in which the company participates.

Let's assume that the competitive battle among companies takes place at five levels.

1. **Quaternary competition.** Among different products which satisfy the same need or want[10]. Product category vs. product category. Different product categories satisfy a need or want in a similar way. This level of competition is the least important competitive threat for the company. It's easy for companies to identify their quarternary competitors quickly. At this level the real competition is among product categories. This level of competition unites the players in a category and they can lead the battle together for the honour (advantages) of their category against other categories and products which satisfy the needs of the consumer in a similar way. This level of competition makes companies share their competitive advantages with their rivals for the benefit of all and of the category. *Janiece Webb*, who heads Motorola's Advanced Technology Businesses, the unit responsible for managing the company's intellectual property portfolio, says: "Taking the traditional protectionist approach to your most valuable technologies doesn't work if you're trying to build a new industry around a common technology platform. We license [the cell-phone communication standard] GSM extensively, for example, and we license it to direct competitors such as Nokia and Ericsson. We want this industry to take off, and giving everyone access to a common technology will help build the industry more quickly, which is good for everyone."[11]
2. **Tertiary competition.** Product line vs. product line of the same product category. GSM Smartphones vs. GSM without OS.
   Different product lines offer different range of benefits and their products have a different 6th level respectively. It's useless to analyse the competitive battle among products which offer different range of benefits. In each case the product line offers more benefits (functions, quality) at a higher price. In these cases the consumer is aware of a clearly established hierarchy of the product lines within one product category. The logic is always one and the same here. Benefits and quality that are highly esteemed by the consumer (image, prestige of the trademark) bear a higher price. These product lines are clearly distinguishable for consumers by their prices.
3. **Secondary competition.** Product type vs. product type of the same product line. Smartphones with Symbian OS vs. Smartphones with other OS. We can assume that the products of the different companies competing in the same product line (product type vs. product type) form strategic product groups by analogy with Porter's idea about strategic groups of companies. The companies which produce products of the same product type form strategic groups according to Porter's criteria.
   Products belonging to a particular product type unite the companies which have produced them in strategic groups of 1st and 2nd type on the basis of a common 6th level. Companies which produce products of a certain product type form a strategic group. Products belonging to the same product type form a strategic product group. Companies from a strategic group produce a strategic product group of the same product type respectively. The 6th product level divides companies into groups producing different product types. Here the competition is much clearer. The competition of a strategic group of companies against another strategic group is one with a clearer focus (foresight). Here the competition is among products of companies belonging to the same product line and the same product category (class), satisfying the same needs and wants and aimed at a certain target segment of consumers on the market. In this case, the competition of product type vs. product type among companies represents the competition of strategic groups of companies vs. strategic groups of companies; all from one group vs. all from the other group; Symbian Smartphones vs. non-Symbian Smartphones.

[10] There is a difference between need and want. Need is the lack of satisfaction or satisfaction that the consumer tries to keep longer in time. Want is a need that has acquired a specific shape in accordance with the economic, social and cultural status of the person.

[11] See: Kline, D., "Sharing the Corporate Crown Jewels", MIT Sloan Management Review 44, no.3 (Spring 2003): 89-93.

4. **Primary competition.** Differentiation and competition among subgroups in strategic groups of first and second type are different.

   **4.1.** Primary differentiation among the brands of the same product type but with a different level of involvement in the strategic group of first type. GSM Smartphones with OS Symbian but with a different level of involvement in the Symbian strategic group. The subgroup of the owners of ISA Symbian Ltd. vs. the subgroup of Symbian OS licensees. We shall call this level of competition and differentiation of subgroup vs. subgroup in the same strategic product group “primary”. Of course, these companies don’t stop competing among themselves although they offer similar products with different trademarks. Here there is no competition at the sixth product level. They compete mainly with relation to the time of innovation launches which are at the other five product levels. As expected, they continue to compete in relation to the other elements of the marketing mix, with relation to the price, distribution and communication with the clients.

   Symbian’s mission is to establish Symbian OS as the world standard for mobile digital data systems and the owners of Symbian Ltd. are successfully achieving their goals. The common sixth level for these two subgroups in the strategic group doesn’t hamper competition but makes it even more interesting which is all for the benefit of the consumers. Differentiation on the basis of the sixth level among the subgroups is impossible because of the equal rights for applying the competitive advantage.

   **4.2** Subgroup vs. subgroup in a strategic group of second type (Microsoft subgroup vs. Linux subgroup). These kinds of strategic groups consist of those rival companies and positions in the market which are divided into strategic groups on the basis of similar product attributes at the sixth product level but of different type and produced by different companies.

   You can see in Figure 5 Microsoft and Linux strategic product subgroups. You probably notice that there are different models of Motorola Smartphones that work with different operating systems (Symbian, Microsoft, and Linux). This allows Motorola to segment the market of prospect consumers in greater detail reinforcing the idea of product mass customization which is possible thanks to the operating systems. The competition between these two operating systems is very much familiar to us. The two most popular operating system for PC in the world confront each other as a sixth level of two different strategic product subgroups. The functions of the sixth product level here are to unite and divide. The compliant product (that’s how we called the sixth level) unites the products of this product type and of this strategic product group. It also unites the companies in a strategic group of second type. The compliant product simultaneously divides them in strategic subgroups on the basis of their differences regarding the different operating systems of each subgroup. In contrast to the differentiation among the subgroups of first type which is difficult to achieve, the differentiation among the subgroups here is easier to be achieved and the groups could differentiate themselves also on the basis of real physical characteristics which is aided by the different ranges of benefits corresponding to the different operating systems. The operating systems are now the real differentiating criteria and the genuine competition is among them. And the battlefield on which this battle is fought is not only our mind but also the different brands of Smartphones in the product subgroups. At this level of competition each product subgroup is united on the basis of a clear set of criteria. The compliant product level is the level of compliance of the product with the given operating system. The battle among different operating systems unites the players in one group and they join their competitive efforts. The primary level of competition is among sub-strategic groups – sub-strategic group vs. sub-strategic group.

5. **Zero competition**. Among the brands of the same strategic subgroup. Level of psychological competition. This is the genuine and the most significant competition for the companies. Brands of the same product type, line and category compete here. This is the level at which the strength (the value and the image) of the brand is of crucial significance because companies compete with almost the same products and product benefits. There is no differentiation on the basis of the sixth level and there is a small, insignificant differentiation on the basis of product attributes at the other product levels. So here the differentiation among brands is mainly done through the strategies of psychological positioning and the place which the product has in the consumer's mind. "Positioning is what you do to the mind of the prospect. That is, you position the product in the mind of the prospect."[12] Psychological positioning is the only thing possible because the products have almost the same technological characteristics and it is very difficult to pinpoint any significant, distinguishable characteristics which are necessary to build a strategy for physical positioning followed by one for psychological positioning.
   Both positioning and differentiation can be psychological. When you can't differentiate yourself on the basis of some real, physical product characteristics the only thing left for you to do is to differentiate yourself psychologically – in the consumer's mind. When you can't achieve real competitive advantages you fabricate them. At this level of competition differentiation is primarily psychological (often substantiated by minor physical differences between brands and models). Here the companies use "the obvious" to differentiate and position themselves. They try to present "the obvious" as their advantage. As an example, UBS Wealth Management, after having been voted "Best Private Bank" by Euromoney, came up with the following advertising motto: "Putting clients first has made us first." But isn't this the motto of every bank in the modern world? Wouldn't every bank say the same? According to the unwritten law, "the obvious" is not a real competitive advantage because it is a characteristic possessed by all competitors. Psychological product differentiation is a differentiation which takes place in the consumer's minds. The companies use either fabricated differentiating advantages or "the obvious" which the consumer has accepted as normal but is presented as a differentiating advantage. Let's consider the following examples of unsuccessful psychological differentiation between Nokia and Samsung:

| Brand: | Motto: |
|---|---|
| Nokia | Connecting people |
| Samsung | Connecting you across the globe |

   "The obvious" is obvious for your competitor, too.

[12] Ries, A., Jack Trout, "Positioning: The Battle for Your Mind," 20th ed. (McGraw0Hill, 2001).

**Figure 7 Merged model of 5 product levels and 5 levels of competition and the possible levels of differentiation at the different levels**

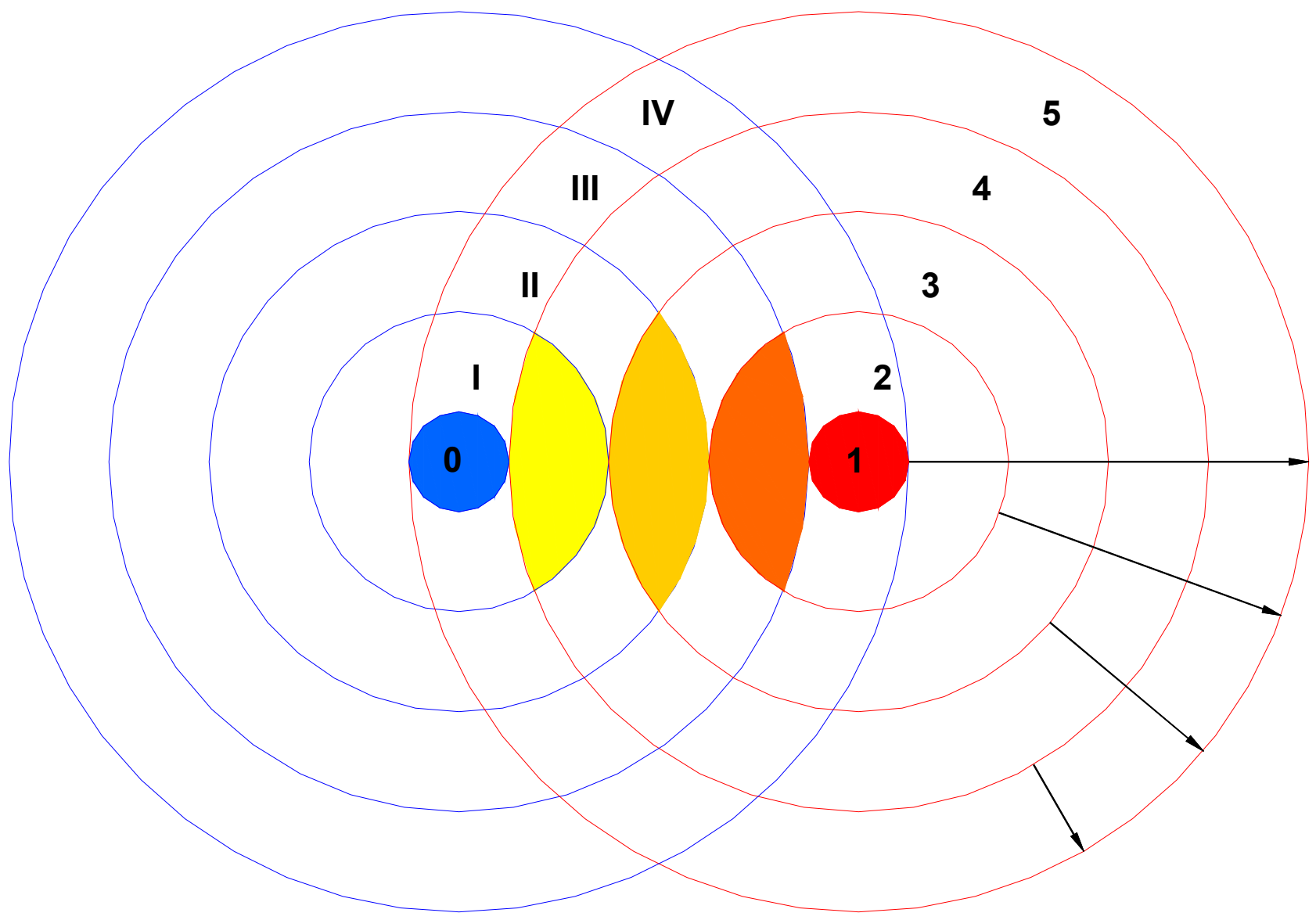

1. Quaternary competition and differentiation. At this level different product categories compete and they satisfy the same needs and wants. Products have a common 1st level and they are free to compete at all others. Here the possibilities for differentiation are the most.
2. Tertiary competition. Products with tertiary competition have the same 1st and 2nd product levels and are free to differentiate and compete among themselves at the other levels.
3. Secondary competition. Products have the same or similar 1st, 2nd and 3rd levels and they distinguish among themselves on the basis of 4th and 5th levels.
4. Primary competition. The only possible differentiation is at the 5th product level.
5. Zero competition. When there is zero competition differentiation is difficult to achieve even at the level of the potential product. One of the few possibilities for differentiation is time of innovation launches or psychological differentiation. It's clear that these innovations are outside the level of compliance with the standard (6th level).

## The sixth product level called compliant product is a connecting element between the physical product characteristics and the strategy of the producer company

Differentiation and competition among products which satisfy the same need or want is done at many levels. Infusing the sixth product level with content is necessary for setting clear standards for the levels of differentiation among the products. Identifying and knowing the proximity of the competitors is the first step towards successful differentiation among them. Differentiation of the products of companies from all other competitors has always been a must for market success. Sometimes companies have to identify their competitors circle and the levels of threat corresponding to their proximity to the company and its products. By determining to which strategic group the company belongs and of which type it is, the company makes the first step towards determining its strengths and weaknesses according to the different levels of competition of each product. The same analysis clearly determines the opportunities and threats resulting from the competitors and the position in the sector.

The meaning and content of the 6th product level and the levels of differentiation and competition help the company to see its competitive position compared to all its competitors in the sector. The model of the levels of differentiation and competition on the basis of the 6th product level can be used as a model for analysis of the marketing environment. It helps the managers to identify "the enemies" and "the friends" of their business. Companies should carefully analyze and evaluate the levels of competition among them. Underestimation of a competitor is as dangerous as overestimation of another. Overestimation of a given competitor, strategic group or subgroup of companies can be very fatal for the business at certain moments since the company puts its competitive efforts in the wrong direction or can lead to wasting unnecessary competitive efforts in many directions due to vagueness about the importance of the threats which would lead to weakening of the competitive strength of the company and would make the company more vulnerable. While I was doing this study and was analyzing the reasons for success and failure of companies in the mobile phone sector, I reached the conclusion that the purposeful participation of a company in strategic groups or strategic subgroups is a must in order to retain one competitive advantage for a longer period of time. When a strategic group is formed on the basis of a common competitive advantage, it immediately stops being one within the group itself. All participants in the group accept the competitive advantage as their own and strive to develop it if possible. Participation in a strategic group or subgroup helps the company to make friends and reduce the number of its enemies. The greatest benefit is for the initiator company which has started and is the first creator of the competitive advantage.

**Motorola's success in the GSM category:** By licensing GSM technology widely, the company created a product category successfully. This, in fact, puts Motorola in the enviable position of essentially taxing its competitors for their help in building the industry.

**Motorola's failure with the satellite-based phone system Iridium:** Morola and other Iridium investors tried to create this category without giving other companies the opportunity to buy licenses for it. They have worked very hard to support Iridium's efforts to reorganize and continue operating the business. Unfortunately, that has not happened. Iridium's spectacular failure, despite more than a decade of planning, is largely the result of poor marketing and a service that was far too costly, particularly as cellular phone coverage and costs improved worldwide, analysts have said. After Iridium's failure in 2000, competitor Globalstar Telecommunications capitalized on Iridium's misfortune by offering rebates to former Iridium users.

**The conclusions**:

- If you want to create a product category, you should create a strategic group of companies around a competitive advantage which should be adopted as a common 6th product level.
- If you want to create a stable, durable competitive advantage, you should create a strategic group of companies which should be united in a product category on the basis of this competitive advantage, adopted by them as a common 6th product level.
- The closer a competitor company is to you on the levels of competition, the better "friend" or the worse "enemy" it could be. And vice versa: the further away it is from you, the weaker "friend" or "enemy" it could be for you.

According to Peter Drucker, each company has to ask itself some seemingly easy questions: "What is our business? Who is the customer? What is the value for the customer? What will our business be? What should our business be?" These are some of the most difficult questions that the company has to answer repeatedly. Apart from them, a contemporary global has to answer one more question: what to create, develop and protect as its competitive advantage and what to share (creating or participating in a strategic group), by selling or buying a license.

* ***Stanimir Andonov*** *is a regular lecturer in Marketing at New Bulgarian University, assistant professor in Corporate Culture and Ethics at the University of National and World Economy and a PhD student at Bulgarian Academy of Sciences. He can be reached at stanimirandonov@nbu.bg, +35988 848 30 89*